\documentclass{aa}                 
\usepackage{graphics}

\newcommand\kms{$\mbox{km\,s}^{-1}$}
\newcommand\bj{$b_{\mathrm J}$}
\newcommand\di{\delta_i}
\newcommand\dsim{\Delta_{\mathrm sim}}
\newcommand\dobs{\Delta_{\mathrm obs}}
\newcommand\hel{{\mathrm hel}}
\newcommand\loc{{\mathrm loc}}
\newcommand\ndec{N_{\mathrm dec}}
\newcommand\ndir{N_{\mathrm dir}}
\newcommand\vmean{\overline{V}}
\newcommand\tablehead{
\begin{flushleft}
\begin{tabular}{rcllllllrrrrr}
\hline
\noalign{\smallskip}
\multicolumn{3}{c}{RA (B1950)} &\ & \multicolumn{3}{c}{Dec (B1950)}
   &\multicolumn{1}{c}{\bj} 
   &\multicolumn{1}{c}{$V^{\mathrm MEF}_{\mathrm ccf}$}
   &\multicolumn{1}{c}{$V^{\mathrm MEF}_{\mathrm emi}$}
   &\multicolumn{1}{c}{$V^{\mathrm OPT}_{\mathrm ccf}$}
   &\multicolumn{1}{c}{$V^{\mathrm OPT}_{\mathrm emi}$}
   &\multicolumn{1}{c}{$V_\hel$}\\
\cline{1-3}\cline{5-7}
\multicolumn{1}{c}{(h)} &\multicolumn{1}{c}{(m)} 
   &\multicolumn{1}{c}{(s)}  & &\multicolumn{1}{c}{(\degr)}
   &\multicolumn{1}{c}{(\arcmin)} &\multicolumn{1}{c}{(\arcsec)} 
   & &\multicolumn{1}{c}{(\kms)} &\multicolumn{1}{c}{(\kms)}
   &\multicolumn{1}{c}{(\kms)} &\multicolumn{1}{c}{(\kms)}
   &\multicolumn{1}{c}{(\kms)}\\
\multicolumn{3}{c}{(1)\phantom{m}} & &\multicolumn{3}{c}{(2)}
   &\multicolumn{1}{c}{(3)}
   &\multicolumn{1}{c}{(4)} &\multicolumn{1}{c}{(5)} &\multicolumn{1}{c}{(6)} 
   &\multicolumn{1}{c}{(7)} &\multicolumn{1}{c}{(8)}\\
\noalign{\smallskip}
\hline
\noalign{\smallskip}
}
\newcommand{\aj}{AJ}         
\newcommand{\aaa}{A\&A}      
\newcommand{\aas}{A\&AS}     
\newcommand{\apj}{ApJ}       
\newcommand{\apjs}{ApJS}     
\newcommand{\mnras}{MNRAS}   

\begin{document}

\thesaurus{20(04.03.1; 11.03.4; 11.04.1)}

\title{Kinematics of the southern galaxy cluster Abell 3733\thanks{Based 
on observations made at the European Southern Observatory, La Silla, Chile.}\fnmsep\thanks{Table 1 is only available in electronic form at the CDS via anonymous ftp to cdsarc.u-strasbg.fr (130.79.128.5) or via http://cdsweb.u-strasbg.fr/Abstract.html}}


\author{Jos\'e M.\,Solanes\inst{1,2} \and Paul Stein\inst{1}}
 \institute{Departament d'Astronomia i Meteorologia, Universitat de
 Barcelona. Av.~Diagonal 647; E--08028~Barcelona, Spain.\and Departament
 d'Enginyeria Inform\`atica, Universitat Rovira i Virgili. Tarragona,
 Spain.\\E-mail: solanes@pcess1.am.ub.es, paul@pcess2.am.ub.es }

\offprints{J.M.\,Solanes}

\date{Received $\ldots$ / Accepted $\ldots$ }

\titlerunning{The cluster Abell 3733}
\authorrunning{J.M.\,Solanes \& P.\,Stein}

\maketitle


\begin{abstract}
We report radial velocities for 99 galaxies with projected positions
within $30\arcmin$ of the center of the cluster A3733 obtained with the
MEFOS multifiber spectrograph at the 3.6-m ESO telescope. These
measurements are combined with 39 redshifts previously published by
Stein (1996) to built a collection of 112 galaxy redshifts in the field
of A3733, which is used to examine the kinematics and structure of this
cluster. We assign cluster membership to 74 galaxies with heliocentric
velocities in the interval $10\,500$--$13\,000$ \kms. From this sample
of cluster members, we infer a heliocentric systemic velocity for A3733
of $11\,653^{+74}_{-76}$ \kms, which implies a mean cosmological
redshift of 0.0380, and a velocity dispersion of $614^{+42}_{-30}$
\kms. The application of statistical substructure tests to a
magnitude-limited subset of the latter sample reveals evidence of
non-Gaussianity in the distribution of ordered velocities in the form
of lighter tails and possible multimodality. Spatial substructure tests
do not find, however, any significant clumpiness in the plane of the
sky, although the existence of subclustering along the line-of-sight
cannot be excluded.
\end{abstract}


\section{Introduction}

The rapid development of multifiber spectroscopy in recent years has
made possible the simultaneous acquisition of large numbers of galaxy
spectra. The obtention of extensive and complete redshift data bases
for clusters of galaxies has hastened the investigation of the physical
properties of their visual component which, in turn, is allowing for a
better understanding of the characteristics of the dark matter
distribution on Mpc scales. Here, we report a total of 104 redshift
measurements for 99 galaxies in the field of A3733 and use these data,
in combination with a previously published sample of 39 redshifts, to
perform a kinematic and spatial analysis of the central regions of this
cluster.

A3733 is a southern galaxy cluster listed in the ACO catalog (Abell,
Corwin, \& Olowin 1989)\cite{ACO89} as of intermediate Abell's
morphological type and richness class $R=1$. This cluster hosts a
central cD galaxy, included in the Wall \& Peacock (1985)\cite{WP85}
all-sky catalog of brightest extragalactic radio sources at 2.7 GHz,
which has led to its classification as of Bautz-Morgan type I--II
(Bautz \& Morgan 1970)\cite{BM70}. A3733 is also one of the 107 nearby
rich ACO clusters ($R\ge 1$, $z\le 0.1$) included by Katgert et
al. (1996)\cite{Ka96} in the ESO Nearby Cluster Survey (ENACS), as well
as a one of the X-ray-brightest Abell clusters detected in the ROSAT
All-Sky Survey by Ebeling et al. (1996)\cite{Eb96}.

The only major kinematical study of A3733 done so far is that of Stein
(1997)\cite{St97}. From a sample of 27 cluster members located within
$r\la 16\arcmin$ from the cluster center, this author has found no
evidence of significant substructure in the cluster core. This study of
A3733, which is part of a more general investigation of the frequency
of substructure in the cluster cores from an optical spectroscopic
survey conducted on a sample of 15 nearby ($0.01\la z\la 0.05$) galaxy
clusters (Stein 1996)\cite{St96}, is based on a dataset that has many
characteristics in common with the ENACS data gathered for the same
field. Indeed, the two datasets have been obtained with the OPTOPUS
multifiber spectrograph at the ESO 3.6-m telescope and cover
essentially the same area on the sky. Besides, they have also a very
similar number of galaxies: 39 and 44, respectively (28 of which are
shared).

The MEFOS redshift dataset for A3733 reported in this paper contains
two and a half times the number of galaxy radial velocities reported by
Stein (1996)\cite{St96}, including 26 reobservations, while it covers a
circular region around the center of A3733 four times
larger. Furthermore, its high degree of completeness offers the
possibility of extracting a complete magnitude-limited subset with a
number of galaxies large enough for its use on statistical
analysis. The plan of the paper is as follows. In Sect.~2 we discuss
the MEFOS spectroscopic observations and data reduction, and present a
final sample with 112 entries built by the combination of the MEFOS and
Stein's (1996)\cite{St96} data. Section~3 begins with a brief
description of the tools which will be used for the analysis of the
data. Next, we identify the galaxies in our sample that belong to
A3733, and use this dataset and a nearly complete magnitude-limited
subset of it to examine the kinematical properties and structure of the
central regions of the cluster. Section~4 summarizes the results of our
study.

\section{MEFOS observations and data reduction}

A total of 104 redshift measurements for 99 galaxies within a circular
region of $30\arcmin$ around the radio position of the cluster cD,
$\mbox{RA} = 20^{\mathrm h} 58^{\mathrm m} 39\fs 0$ and $\mbox{Dec} =
-28\degr 15\arcmin 22\arcsec$ (Tadhunter et al. 1993)\cite{Ta93}, were
obtained using the MEFOS multifiber spectrograph at the 3.6-m ESO
telescope at La Silla (Chile). The observations were carried out on May
23--26, 1995 during an observing run whose main target was the dwarf
galaxy population of the Centaurus cluster (see Stein et
al. 1997)\cite{SJF97}. The MEFOS instrument has a circular field of
view of $1\degr$ and 29 fiber arms which carry two spectral fibers of
$2\farcs 5$ aperture for simultaneous object and sky acquisition, and
one image fiber of $36\arcsec\times 36\arcsec$ for the interactive
repositioning of the spectral fibers. A grating with 300
lines~mm$^{-1}$ was used to produce spectra in the range between 
3800 and 6100~\AA\ with a typical resolution of ca. 10~\AA. The detector
was a TI $512\times 512$ CCD chip.

The raw CCD spectra were reduced using the MIDAS package through
several steps which include cleaning from defects, cosmic ray removal,
flat-fielding, one-dimensional extraction, and wavelength calibration
using a He-Ne lamp before and after each exposure. The sky subtraction
was performed subtracting from each one of the object spectra the mean
of the output of all the fibers positioned on blank sky positions from
the corresponding exposure. Prior to sky subtraction the signal of each
spectrum (including the sky spectra) was scaled with respect to the
intrinsic transmission efficiency of the corresponding fiber, which had
been determined using the average over the observed fields of the signal
under the O\,{\sc i} emission line at 5577.4 \AA. 

After the final one-dimensional spectra had been extracted, velocities
were computed either from emission lines or from absorption lines, or
from both. Emission-line redshifts were obtained from galaxies with at
least two clearly visible emission lines (mostly O\,{\sc ii},
H{$\beta$}, and O\,{\sc iii}). Their redshifted positions were
determined from fits with a Gaussian superposed onto a quadratic
polynomial approximating the local continuum. The final redshift of a
galaxy was then computed as the unweighted mean over the $n$
emission lines present in its spectrum. Since the errors in the
redshift measurement of each single line are essentially dominated by
uncertainties in the wavelength calibration (Stein
1996)\cite{St96}, individual measurement errors were taken equal to 100
\kms, independently of line strength. Accordingly, an uncertainty of
$100/\sqrt{n}$ \kms\ was assigned to emission-line 
redshifts. Absorption-line redshifts were obtained using the
standard cross-correlation algorithm described by Tonry \& Davis
(1979)\cite{TD79}. This technique requires the previous removal of both
galaxy emission lines and strong night-sky lines, and the
transformation of the spectral continuum to a constant level of
zero. Special care was taken that the continuum subtraction did not
create spurious features of low spatial frequency which could be
confused with broad, superposed absorption lines. For the determination
of the zero-point shift one single template was constructed by merging
20 galaxy spectra with high S/N and well known redshifts. Only
normalized cross-correlation peaks of height 0.25 or larger were
considered as significant. Both emission-line and cross-correlation
redshifts were then corrected to heliocentric values.

The 26 galaxies observed also by Stein (1996)\cite{St96} in the field
of A3733 with the OPTOPUS spectrograph were used to determine the
scaling factor of the internal errors estimated in the
cross-correlation procedure, resulting in external errors of typically
40--50 \kms. These same galaxies gave a mean velocity difference of
$-23$ \kms\ (the MEFOS redshifts being typically smaller), consistent
with zero to within the reported measurement errors. Only for 5 of our
galaxies we could measure both emission-line and cross-correlation
redshifts. Again, an excellent consistency was found between the two
kinds of measurements.

The cross-correlation and emission-line radial velocities
for the 99 galaxies observed with the MEFOS spectrograph in the field
of A3733 are listed in Cols. (4) and (5) of Table 1, together with
their estimated external errors. Columns (6) and (7) give the same
information for the 39 galaxies observed with the OPTOPUS instrument by
Stein (1996)\cite{St96}. Column (8) lists the final radial velocities
and their estimated uncertainties which result form a weighted average
of the data in Cols.~(4)--(7). The combination of these two samples
gives a total of 112 entries, which will be used in the following
section to examine the kinematical properties and structure of
A3733. This is about three times the number of galaxies used in the
previous study by Stein (1997)\cite{St97}. The first three columns of
Table 1 contain the celestial coordinates for the epoch B1950.0 and the
\bj\ magnitudes from the COSMOS catalog kindly provided by
H. MacGillivray. The completeness in apparent magnitude of the final
dataset is high, with percentages of 100, 92, 75, 63, and 50\% of all
known (COSMOS) galaxies in the same region of the sky covered by our
observations at the \bj\ magnitude limits of 17.0, 17.5, 18.0, 18.5,
and 19.0, respectively.

\section{Kinematic and spatial analysis}

Following the recommendations of Beers, Flynn, \& Gebhardt
(1990)\cite{BFG90}, we will characterize the velocity distribution of
our cluster sample by means of the biweight estimators of central
location (i.e., systemic velocity), $\vmean$, and scale (i.e., velocity
dispersion), $\sigma$. We will assign errors to these estimates equal
to the 68\% bias-corrected bootstrap confidence intervals inferred from
$10\,000$ resamplings. The program ROSTAT, kindly provided by T. Beers,
will be used for all these calculations.

The ROSTAT program includes also a wide variety of statistical tests,
which can be used to assess the consistency of the empirical
line-of-sight velocity distribution of the A3733 members (see next
subsection) with draws from a single Gaussian parent population. A fair
representation of the overall results of the ROSTAT tests will be given
by quoting the value of the statistic and associated probability for
the canonical $B_1$ and $B_2$ tests, which measure, respectively, the
skewness (asymmetry) and curtosis (elongation) of the velocity
distribution, and for the Anderson-Darling $A^2$ omnibus
test. Definitions of these tests can be found in Yahil \& Vidal
(1977)\cite{YV77} and D'Agostino (1986)\cite{Da86}. The Gaussianity
tests will be complemented by the Dip test of Hartigan \& Hartigan
(1985)\cite{HH85}, which tests the hypothesis that a sample is drawn
from a unimodal (though not necessarily Gaussian) parent distribution,
and by the search of individual weighted gaps, $g_\ast$, in the
velocity distribution of size 2.75 or larger (for a definition of
weighted gap see, for instance, Beers et
al. 1990)\cite{BFG90}. Individual weighted gaps this large are highly
significant since they arise less than 1\% of the time in random draws
from a Gaussian distribution, independently of sample size. We refer
the reader to the listed sources and references therein for a detailed
explanation of these statistical techniques.

We will investigate also the presence of substructure in the spatial
distribution of galaxies by means of two powerful tests. First, we will
apply a 2D test developed by Salvador-Sol\'e, Sanrom\'a, \&
Gonz\'alez-Casado (1993)\cite{SSG93}, hereafter referred to as the SSG
test, which relies exclusively on the projected positions of galaxies
on the sky (though velocity information is required to define strict
cluster membership). This test produces two different estimates of the
projected number density profile of the cluster, $\ndec(r)$ and
$\ndir(r)$, which are, respectively, sensitive and insensitive to the
existence of correlation in the galaxy positions relative to the
cluster background density. The subscript ``dec'' identifies the
density profile obtained via the \emph{deconvolution} of the histogram
of intergalaxy separations, while the subscript ``dir'' applies to the
density profile arising \emph{directly} from the integral of the
histogram of clustercentric distances of the cluster galaxies (eqs.~[4]
and [6], respectively, in Salvador-Sol\'e et al. 1993\cite{SSG93}). The
two profiles are convolved with a window of smoothing size
$\lambda_{\mathrm min}$ corresponding to the minimum resolution-length
imposed by the calculation of $\ndec(r)$. The significance of
substructure is estimated from the null hypothesis that $\ndec(r)$
arises from a Poissonian realization of some (unknown) theoretical
density profile which has led to the observed radial distribution of
galaxies. The probability of this being the case is calculated by means
of the statistic:
\begin{equation}\label{chi2}  
\chi^2 = {\left(\ndec(0)-\ndir(0)\right)^2 \over{2S^2(0)}}\;,
\end{equation} 
for one degree of freedom. In eq.~(\ref{chi2}), $\ndec(0)$ and
$\ndir(0)$ are the central values of the respective density profiles of
the cluster, while $S^2(0)$ is the central value of the radial run of
the variance of $\ndir(r)$ calculated from a set of simulated clusters
convolved to the $\lambda_{\mathrm min}$ imposed by $\ndec(r)$. The
simulated clusters are generated by the azimuthal scrambling of the
observed galaxy positions around the center of the cluster, i.e., by
randomly shuffling between 0 and $2\pi$ the azimuthal angle of each
galaxy, while maintaining its clustercentric distance unchanged. It
must be stressed, however, that the sensitivity of the SSG test is not
affected by deviations of the spatial distribution of the galaxy sample
under scrutiny from circular symmetry (see Salvador-Sol\'e et
al. 1993\cite{SSG93}). It is also worth noting that this test does not
require a priori assumptions on the form of the true projected number
density profile of the cluster, nor on the number and size of the
subgroups that might be present in the data.

\begin{figure}
 \resizebox{\hsize}{!}{\includegraphics{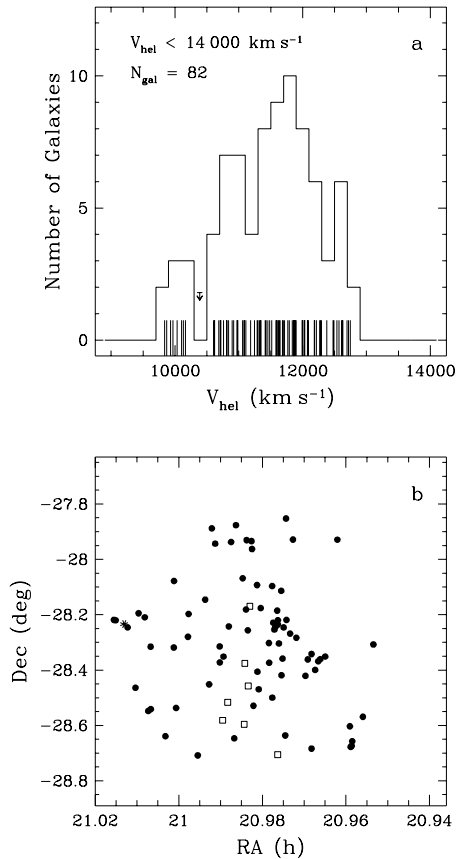}} 
  \caption{{\bf a} Stripe density plot and velocity histogram of the
  galaxies with $V_\hel< 14\,000$ \kms\,in the A3733 sample. The arrow
  marks the location of a highly significant weighted gap ($p=0.001$)
  in the velocity distribution. {\bf b} Corresponding spatial
  distribution. The 7 members of a suspect foreground group are
  identified by open squares, while the asterisk marks the galaxy with
  the lowest $V_\hel$. Filled circles identify our choice of strict
  cluster members.}
  \label{Fig. 1}
\end{figure}

The second spatial substructure test that will be applied to our data
is the 3D Dressler \& Shectman (1988b; DS)\cite{DS88b} test, which is
sensitive to local kinematic deviations in the projected galaxy spatial
distribution. The DS test assigns a local estimate of the velocity
mean, $\vmean_\loc$, and dispersion, $\sigma_\loc$, to each galaxy with
a measured radial velocity. These values are then compared with the
values of the kinematical parameters for the entire sample. The
statistic used to quantify the presence of substructure is the sum of
the local kinematic deviations for each galaxy, $\di$, over the $N$
cluster members, which we will calculate through the expression:
\begin{eqnarray} \label{delta}  
\Delta & = & \sum_{i=1}^N\di \nonumber \\
& = & \sum_{i=1}^N\left[{N_{\mathrm
kern}+1\over{\sigma^2}}\left(\left(\vmean_{\loc,i}-\vmean\right)^2+\left(\sigma_{\loc,i}-\sigma\right)^2\right)\right]^{1/2}\;,
\end{eqnarray}
where $N_{\mathrm kern}={\mathrm nint}(\sqrt N)$, ${\mathrm nint}(x)$
stands for the integer nearest to $x$, maximizes the sensitivity of
the DS test to significant substructure (Bird 1994).\cite{Bi94} To
avoid the formulation of any hypothesis on the form of the velocity
distribution of the parent population, the DS test calibrates the
$\Delta$ statistic by means of Monte-Carlo simulations that randomly
shuffle the velocities of the galaxies while keeping their observed
positions fixed. In this way any existing correlation between
velocities and positions is destroyed. The probability of the null
hypothesis that there are no local correlations between the position
and velocity of the cluster members is given in terms of the fraction
of simulated clusters for which their cumulative deviation, $\dsim$, is
smaller than the observed value, $\dobs$. Again, we refer the reader to
the quoted references for further details on the two spatial
substructure tests used in the present analysis.

\subsection{The sample of cluster members}

Before we can investigate the presence of substructure in A3733 we need
to assign cluster membership to the galaxies in our sample. Examination
of the radial velocities of the 112 galaxies listed in Table~1 allows
the exclusion of 30 obvious interlopers (all background galaxies and
groups), which are separated by more than 6500 \kms\ from the main
velocity group. Subsequent membership assignment for the remaining 82
galaxies is based on the their velocity distribution and projected
positions, displayed in Figs.~1a and 1b, respectively. These figures
show the existence of 8 objects with velocities smaller than $10\,500$
\kms\ separated from the other galaxies by a gap in heliocentric
velocity of $\sim 450$ \kms. Seven of these galaxies appear also to be
concentrated on a small area of the sky. The cluster diagnostics
described at the beginning of this section reveal that the above gap in
velocity corresponds to an individually large normalized gap of size
3.39 in the 82 ordered velocities. The ``per-gap'' probability for a
weighted gap this size is only 0.001. This and the fact that the
suspected foreground group of 7 galaxies has a velocity dispersion of
only 73 \kms\ suggest that it might constitute a separate dynamical
entity. Accordingly, we chose to consider {\it bona fide\/} A3733
members the 74 galaxies in our sample with heliocentric velocities
between $10\,500$ and $13\,000$ \kms. Note that we are excluding also
from cluster membership the remaining foreground object with the lowest
measured radial velocity. From the set of cluster members, we obtain
$\vmean_\hel=11\,653^{+74}_{-76}$ \kms\ and $\sigma =614^{+42}_{-30}$
\kms\ after applying relativistic and measurement error corrections
(Danese, De Zotti, \& di Tullio 1980)\cite{DDD80}. These values are
compatible, within the adopted uncertainties, with the values
$\vmean_\hel=11\,716\pm 103$ \kms\ and $\sigma =522\pm 84$ \kms\
obtained in the previous analysis of this cluster by Stein
(1997)\cite{St97} from a sample containing 27 of the current cluster
members. The mean heliocentric velocity calculated for A3733 results in
a mean cosmological redshift of $\,\overline{z}_{\mathrm CMB}=0.0380$
after correction to the CMB rest frame (Kogut et
al. 1993)\cite{Ko93}. At the cosmological distance of A3733, one Abell
radius, $r_{\mathrm A}$ ($\equiv 1.5\; h^{-1}$ Mpc), is equal to 0.805
degrees. The subset of 82 galaxies with $V_\hel<13\,000$ \kms\ has
$\vmean_\hel=11\,532^{+94}_{-89}$ \kms, $\sigma =754^{+65}_{-48}$ \kms,
$\overline{z}_{\mathrm CMB}=0.0385$, and $r_{\mathrm A}=0.812$
degrees.

The values of the kinematical parameters of the cluster have been
calculated without taking into account its dynamical state. Indeed, the
visual inspection of Fig.~1a yields suggestive indication of deviation
of the velocity distribution from Gaussianity in the form of lighter
tails and a hint of multimodality. The dictum of the $B_2$ statistic,
which indicates the amount of elongation in a sample relative to the
Gaussian, confirms the platycurtic behavior (i.e., $B_2<3$) of the
velocity histogram giving only a $0.001$ probability that it could have
arisen by chance from a parent Gaussian population. Nevertheless, the
results of the $B_1$ and $A^2$ tests do not indicate significant
departures from normality. As for the possible multimodality, the Dip
test cannot reject the unimodal hypothesis, nor we detect the presence
of highly significant large gaps in the ordered velocities.

Comparable results are obtained if we remove from the sample of cluster
members those galaxies with strong emission lines in their
spectrum. Indeed, the spatial distribution and kinematic properties of
these latter galaxies are similar to those of the galaxies for which
only cross-correlation redshifts are available. Specifically, for the
12 cluster members with emission-line redshifts we find
$\vmean_\hel=11\,416^{+256}_{-218}$ \kms\ and $\sigma
=694^{+164}_{-101}$ \kms, while the remaining 62 galaxies have
$\vmean_\hel=11\,694^{+80}_{-83}$ \kms\ and $\sigma =594^{+44}_{-37}$
\kms.

\subsection{The magnitude-limited sample}

In order to mitigate the effects of incomplete sampling which may
contaminate the results of the statistical tests, especially of those
relying on local spatial information, we concentrate our subsequent
analysis on the subset of 37 members of A3733 with \bj\ $\leq 18$, for
which our original redshift sample contains 75\% of the COSMOS
galaxies. This magnitude limit is chosen as a compromise between
defining a sample (nearly) free of sampling biases and simultaneously
having a large enough number of objects for the detection of
substructure not to be affected by Poissonian errors.

For this sample, the Gaussianity tests confirm essentially the results
obtained for the whole set of cluster members: the $B_2$ test rejects
the Gaussian hypothesis at the 6\% significance level, while the $B_1$
and $A^2$ tests are consistent with a parent normal
population. Remarkably, the results of the other two 1D tests are now
substantially different: the Dip test rejects the hypothesis of
unimodality at the 4\% significance level, while a large gap of size
roughly 230 \kms\ ($g_\ast=3.14$, $p=0.002$) appears near the middle of
the distribution ($V_\hel\sim 11\,500$ \kms) of velocities.

The kinematical complexity of the inner regions of A3733 suggested by
these latter results is not reflected, however, on the spatial
distribution of the galaxies. The SSG test gives, for 1000 realizations
of the cluster generated by the azimuthal scrambling of the galaxy
positions around the location of the cD (see Sect. 2), a 56\% probability
that there is no substructure, which is nonsignificant. The resulting
$\lambda_{\mathrm min}$ of $16\farcm 7$ ($\equiv 0.52\,h^{-1}$~Mpc)
puts an upper limit to the half-coherence length of any possible clump
that may remain undetected in the central regions of A3733. This value
is above the typical scale-length of $\sim 0.3\,h^{-1}$~Mpc of the
clumps detected by Salvador-Sol\'e, Gonz\'alez-Casado, \& Solanes
(1993)\cite{SGS93} in the Dressler \& Shectman's (1988a)\cite{DS88a}
clusters. This suggests that the presence of significant substructure
in the magnitude-limited sample might be hidden by the large smoothing
scale imposed by the calculation of $\ndec(r)$. We have investigated
this possibility by applying also the SSG test to the sample containing
all the 74 cluster members, for which the minimum resolution-length
reduces to only $3\farcm 69$ ($\equiv 0.12\,h^{-1}$~Mpc). In spite of
the fact that this latter sample is biased towards the most populated
regions of A3733, therefore emphasizing any possible clumpiness of the
galaxy distribution on the plane of the sky, we still obtain a 14\%
probability for the null hypothesis.

\begin{figure*}
 \resizebox{\hsize}{!}{\includegraphics{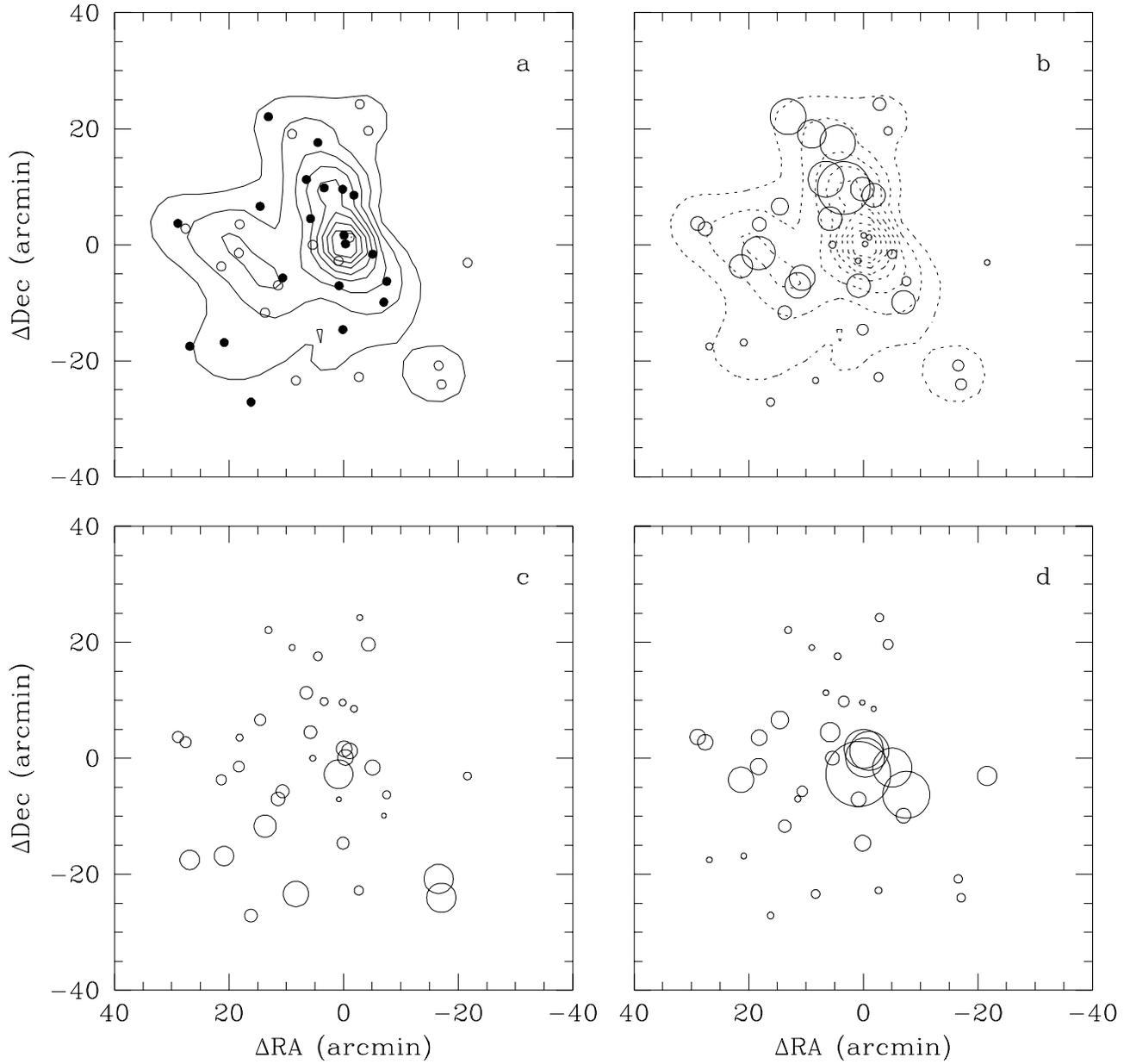}} 
  \caption{{\bf a} Spatial distribution of the 37 galaxies belonging
  to the magnitude-limited sample (\bj\,$\le 18$) of A3733
  members. Galaxies with $V_\hel\leq 11\,500$ \kms\ are identified by
  empty circles, while solid circles mark the location of the galaxies
  with $V_\hel > 11\,500$ \kms. Curves are equally spaced contours of
  the adaptive kernel density contour map for this sample. The contours
  range from $2.18\times 10^{-4}$ to $1.88\times 10^{-3}
  \mbox{\,galaxies\,arcmin}^{-2}$. The initial smoothing scale is set
  to $12\farcm 6$. {\bf b} Local deviations from the global kinematics
  as measured by the DS test. Open circles drawn at the position of the
  individual galaxies scale with the deviation of the local kinematics
  from the global kinematics, $\di$, from which the test statistic
  $\dobs=\sum\di$'s is calculated (see text). The adaptive kernel
  contour map is superposed (dashed lines). {\bf c and d} Monte-Carlo
  models of the magnitude-limited sample obtained after 1000 random
  shufflings of the observed velocities: {\bf c} model with the
  cumulative deviation $\dsim$ closest to the median of the
  simulations; {\bf d} model whose $\dsim$ is closest to the value of
  the upper quartile. Spatial coordinates are relative to the cluster
  center (see text).}
  \label{Fig. 2}
\end{figure*}

The DS test also points to the lack of significant substructure in the
magnitude-limited sample: more than 15\% of the values of the statistic
$\dsim$ obtained in 1000 Monte-Carlo simulations of this sample are
larger than $\dobs$. A visual judgment of the statistical significance
of the local kinematical deviations can be done by comparing the plots
in Figs.~2a--d. Figure~2a shows the spatial distribution of the
galaxies superposed on their adaptive kernel density contour map (see
Beers 1992\cite{Be92} and references therein for a description of the
adaptive kernel technique). The primary clump in this map is centered
at the position of the cD galaxy and is elongated along the north-south
axis; a mild density enhancement can be seen at the plot coordinates
(17,-3). In this figure galaxies with $V_\hel\leq 11\,500$ \kms\ are
represented by empty circles, while solid circles mark the location of
those with $V_\hel > 11\,500$ \kms. Although there is no strong spatial
segregation among the galaxies belonging to each of these two velocity
subgroups, the galaxies included in the second one dominate the central
density enhancement. In Fig.~2b each galaxy is identified with a circle
whose radius is proportional to $\mathrm{\exp(\di)}$, where $\di$ is
given by eq.~(\ref{delta}). Hence, the larger the circle, the larger
the deviation from the global values (but beware of the insensitivity
of the $\di$'s to the sign of the deviations from the mean cluster
velocity). The superposition of the projected density contours shows
that most of the galaxies to the north of the density peak, and to a
lesser extent those closest to the center of the eastern small density
enhancement, have apparently large local deviations from the global
kinematics. The remaining figures show two of the 1000 Monte-Carlo
models performed: Fig.~2c corresponds to the model whose $\dsim$ is
closest to the median of the simulations, while Fig.~2d corresponds to
the model with a $\dsim$ closest to the value of the upper
quartile. The comparison of Fig.~2b with these last two figures shows
that the observed local kinematical deviations are indeed not
significant.

As commented in the Introduction, Stein (1997)\cite{St97} has not found
either any evidence of significant clumpiness on his A3733 OPTOPUS data
(see his Table~3). Nevertheless, we caution that this previous study is
restricted to the innermost ($r\le 16\arcmin$) regions of the cluster
and that it uses, due to the small size of the sample, all the redshifts
available without regard to their completeness.

The results of all the statistical tests applied to our
magnitude-limited sample are summarized in Table~2, together with the
results obtained from the whole sample of cluster members, for
comparison. In Col.~(1) we list the name of the sample and in Col.~(2)
the number of galaxies in it. Columns (3)--(14) give the values of the
test statistic and associated significance levels for the $B_1$, $B_2$,
$A^2$, Dip, SSG, and DS tests, respectively. The significance levels
refer to the probability that the empirical value of a given statistic
could have arisen by chance from the null hypothesis. Thus, the smaller
the quoted probability the more significant is the departure from it.

\setcounter{table}{1}
\begin{table*}
\caption[]{Results of the statistical tests}
\begin{flushleft}
\begin{tabular}{ccccclccccccll}
\hline
\noalign{\smallskip}
\multicolumn{1}{c}{Sample} & \multicolumn{1}{c}{$N_{\mathrm gal}$}
   & \multicolumn{1}{c}{$B_1$}
   & \multicolumn{1}{c}{$p(B_1)$} & \multicolumn{1}{c}{$B_2$} 
   & \multicolumn{1}{c}{$p(B_2)$} & \multicolumn{1}{c}{$A^2$}
   & \multicolumn{1}{c}{$p(A^2)$} & \multicolumn{1}{c}{Dip} 
   & \multicolumn{1}{c}{$p({\mathrm Dip})$} & \multicolumn{1}{c}{$\chi^2$}
   & \multicolumn{1}{c}{$p(\chi^2)$} 
   & \multicolumn{1}{c}{$\dobs$}
   & \multicolumn{1}{c}{$p(\dobs)$}\\
\multicolumn{1}{c}{(1)} &\multicolumn{1}{c}{(2)} &\multicolumn{1}{c}{(3)}
   &\multicolumn{1}{c}{(4)} &\multicolumn{1}{c}{(5)} &\multicolumn{1}{c}{(6)} 
   &\multicolumn{1}{c}{(7)} &\multicolumn{1}{c}{(8)} &\multicolumn{1}{c}{(9)}
   &\multicolumn{1}{c}{(10)} &\multicolumn{1}{c}{(11)} 
   &\multicolumn{1}{c}{(12)} &\multicolumn{1}{c}{(13)}
   &\multicolumn{1}{c}{(14)}\\
\noalign{\smallskip}
\hline
\noalign{\smallskip}
Mag.-limited &37&0.20&0.28&2.08&0.06&0.52&0.18
   &0.08&0.04&0.35&0.56&40.8&0.15\\
All members &74&0.00&0.50&1.99&0.001&0.51&0.20
   &0.03&0.99&2.19&0.14&107.&0.003\\ 
\noalign{\smallskip}
\end{tabular}
\end{flushleft}
\label{Table2}
\end{table*}

\section{Summary}

We have reported 104 radial velocity measurements performed with the
MEFOS multifiber spectrograph at the 3.6-m ESO telescope for 99
galaxies in a region of $30\arcmin$ around the center of the cluster
A3733. To augment this data, we have combined the MEFOS measurements
with 39 redshifts measured by Stein (1996)\cite{St96} with the OPTOPUS
instrument at the same telescope. This has given a final dataset with a
total of 112 entries in the field of A3733. Radial velocities have been
then supplemented by COSMOS \bj\ magnitudes and accurate sky positions
in order to investigate the kinematics and structure of the central
regions of the cluster.

From a sample containing 74 strict cluster members, we have derived a
heliocentric systemic velocity for A3733 of $11\,653^{+74}_{-76}$ \kms,
resulting in a $\overline{z}_{\mathrm CMB}$ of 0.0380, and a velocity
dispersion of $614^{+42}_{-30}$ \kms, in good agreement with the
estimates by Stein (1997)\cite{St97} from the OPTOPUS data
alone. Statistical tests relying exclusively on the distribution of
observed velocities have yield suggestive indication of the possible
kinematical complexity of A3733, especially when applied to a nearly
complete magnitude-limited (\bj\ $\leq 18$) sample of cluster
members. Despite this result, two powerful substructure tests that
incorporate spatial information have failed to detect in this latter
sample any statistically significant evidence of clumpiness in the
galaxy component, in agreement with the findings of a previous study
based on a spatially less extended and less complete dataset. Given
that the sensitivity of the spatial substructure tests we have used is
reduced when the subunits are seen with small projected separations,
the results of the present study cannot exclude, however, the
possibility that the signs of kinematical complexity detected in the
velocity histogram of A3733 might be due to the existence of
galaxy subcondensations superposed along the line-of-sight.

\begin{acknowledgements}
This work has been supported by the Direcci\'on General de
Investigaci\'on Cient\'{\i}fica y T\'ecnica, under contract
PB96--0173. P.S. acknowledges partial support from a research network
grant by the Commission of the European Communities.
\end{acknowledgements}

\newpage
\renewcommand{\baselinestretch}{1}

\begin{table*}
\caption[]{MEFOS and OPTOPUS multifiber spectra velocities for A3733}
\tablehead
 20&56&46.25& &$-28$&27&37.4&18.85&39200$\pm$96& &             &             &            39200$\pm$96\\
 20&56&55.12& &$-28$&20&44.5&18.85& &            29088$\pm$58& &             &            29088$\pm$58\\
 20&56&59.68& &$-28$&19&25.5&18.43&39708$\pm$61& &             &             &            39708$\pm$61\\
 20&57&03.80& &$-28$&29&25.4&17.40&29084$\pm$116& &             &             &            29084$\pm$116\\
 20&57&08.92& &$-28$&16&19.8&16.68&28920$\pm$83&28862$\pm$71& &             &            28886$\pm$54\\
 20&57&12.38& &$-28$&18&27.0&17.81&10840$\pm$83& &             &             &            10840$\pm$83\\
 20&57&14.91& &$-28$&39&56.3&19.07&48045$\pm$119& &             &             &            48045$\pm$119\\
 20&57&21.44& &$-28$&34&07.4&18.45&12268$\pm$82& &             &             &            12268$\pm$82\\
 20&57&27.01& &$-28$&07&37.3&18.50&49400$\pm$80& &             &             &            49400$\pm$80\\
 20&57&30.62& &$-28$&39&26.5&17.89&10710$\pm$46& &             &             &            10710$\pm$46\\
 20&57&31.04& &$-28$&40&26.2&18.38&12086$\pm$50& &             &             &            12086$\pm$50\\
 20&57&31.95& &$-28$&40&39.6&18.57&12078$\pm$58& &             &             &            12078$\pm$58\\
 20&57&32.63& &$-28$&36&10.9&17.49&11348$\pm$42& &             &             &            11348$\pm$42\\
 20&57&35.15& &$-28$&11&42.8&18.97& &            31935$\pm$50& &             &            31935$\pm$50\\
 20&57&43.41& &$-27$&55&45.1&18.09&11518$\pm$101& &             &             &            11518$\pm$101\\
 20&57&53.77& &$-28$&35&02.3&19.27&55141$\pm$78& &             &             &            55141$\pm$78\\
 20&57&53.91& &$-28$&21&04.7&18.12&12606$\pm$41& &            12619$\pm$23& &            12614$\pm$27\\
 20&57&58.20& &$-28$&21&38.6&     & &             &             &            12073$\pm$71&12073$\pm$71\\
 20&57&59.91& &$-28$&22&04.8&     & &             &             &            12615$\pm$71&12615$\pm$71\\
 20&58&01.30& &$-28$&42&44.4&17.80&22274$\pm$50& &             &             &            22274$\pm$50\\
 20&58&01.59& &$-27$&54&04.9&17.89&29051$\pm$163& &             &             &            29051$\pm$163\\
 20&58&02.56& &$-28$&23&57.5&18.21&11991$\pm$62& &             &             &            11991$\pm$62\\
 20&58&05.64& &$-28$&41&01.5&18.19&11329$\pm$39& &             &             &            11329$\pm$39\\
 20&58&05.66& &$-28$&20&30.8&18.95& &             &             &            11682$\pm$71&11682$\pm$71\\
 20&58&08.87& &$-28$&21&41.1&17.74&11593$\pm$46& &            11580$\pm$23& &            11585$\pm$28\\
 20&58&10.91& &$-28$&25&15.8&17.36&11777$\pm$41& &            11954$\pm$22& &            11885$\pm$26\\
 20&58&13.52& &$-28$&19&10.0&17.86&22333$\pm$152& &             &             &            22333$\pm$152\\
 20&58&13.99& &$-28$&12&59.2&     & &             &             &            22623$\pm$71&22623$\pm$71\\
 20&58&16.56& &$-28$&23&36.2&17.74&21889$\pm$44& &             &             &            21889$\pm$44\\
 20&58&18.73& &$-28$&16&59.4&17.69&12087$\pm$40& &            11939$\pm$31& &            12025$\pm$30\\
 20&58&20.41& &$-28$&16&38.2&     & &             &             &            27057$\pm$58&27057$\pm$58\\
 20&58&21.60& &$-28$&23&59.4&18.66&21603$\pm$73& &            21985$\pm$25& &            21945$\pm$34\\
 20&58&21.66& &$-27$&55&42.7&16.87&10759$\pm$41& &             &             &            10759$\pm$41\\
 20&58&24.11& &$-28$&16&05.7&18.34& &            11302$\pm$58& &             &            11302$\pm$58\\
 20&58&26.27& &$-28$&34&26.9&19.06&51176$\pm$171& &             &             &            51176$\pm$171\\
 20&58&27.24& &$-28$&13&08.7&18.17&12562$\pm$53& &            12568$\pm$19& &            12567$\pm$25\\
 20&58&27.64& &$-27$&51&07.8&17.59&10688$\pm$43& &             &             &            10688$\pm$43\\
 20&58&28.30& &$-28$&38&10.2&17.24&11244$\pm$50& &             &             &            11244$\pm$50\\
 20&58&29.69& &$-28$&14&44.4&     & &             &             &            11457$\pm$71&11457$\pm$71\\
 20&58&30.64& &$-28$&21&31.3&18.82&12188$\pm$105& &             &             &            12188$\pm$105\\
 20&58&31.54& &$-28$&25&08.0&     & &             &            10921$\pm$20& &            10921$\pm$30\\
 20&58&31.78& &$-28$&06&49.0&17.00&11771$\pm$35& &            11765$\pm$16& &            11767$\pm$20\\
 20&58&33.75& &$-28$&18&14.6&     & &             &            11896$\pm$16& &            11896$\pm$24\\
 20&58&34.62& &$-28$&13&11.1&     & &             &            12380$\pm$23& &            12380$\pm$35\\
 20&58&34.69& &$-28$&14&07.7&17.49&10820$\pm$46& &            10804$\pm$13& &            10807$\pm$18\\
 20&58&34.81& &$-28$&42&19.0&17.59&10103$\pm$59& &             &             &            10103$\pm$59\\
 20&58&35.24& &$-28$&11&06.6&18.92&12033$\pm$101& &             &             &            12033$\pm$101\\
 20&58&36.94& &$-28$&14&42.9&     & &             &            11632$\pm$30& &            11632$\pm$45\\
 20&58&37.67& &$-28$&15&13.3&16.36&12217$\pm$47& &            12215$\pm$16& &            12215$\pm$21\\
 20&58&38.66& &$-28$&13&44.1&15.07&11875$\pm$44& &            11865$\pm$17& &            11868$\pm$22\\
 20&58&39.47& &$-28$&29&58.6&16.43&11935$\pm$46& &            11861$\pm$16& &            11877$\pm$21\\
 20&58&39.60& &$-28$&05&47.3&16.47&11641$\pm$44& &            11645$\pm$15& &            11644$\pm$20\\
\end{tabular}
\end{flushleft}
\label{Table1}
\end{table*}

\setcounter{table}{0}
\begin{table*}
\caption[]{(cont.)}
\tablehead
 20&58&42.15& &$-28$&22&24.9&16.47&11822$\pm$39& &            11870$\pm$17& &            11855$\pm$22\\
 20&58&42.39& &$-28$&18&08.4&17.87&11032$\pm$38& &            11114$\pm$20& &            11083$\pm$24\\
 20&58&44.37& &$-28$&00&14.1&17.66&22102$\pm$81&22052$\pm$50& &             &            22066$\pm$42\\
 20&58&45.08& &$-28$&11&59.0&     & &             &            28821$\pm$42& &            28821$\pm$63\\
 20&58&49.42& &$-28$&10&32.8&18.72&11585$\pm$52& &             &             &            11585$\pm$52\\
 20&58&51.15& &$-28$&28&10.2&18.48& &            10978$\pm$50& &             &            10978$\pm$50\\
 20&58&52.27& &$-28$&24&21.0&     & &             &            11429$\pm$28& &            11429$\pm$42\\
 20&58&52.58& &$-28$&05&33.8&17.69&12748$\pm$103& &             &             &            12748$\pm$103\\
 20&58&55.68& &$-28$&31&44.5&18.19&11438$\pm$44& &            11512$\pm$24& &            11482$\pm$28\\
 20&58&56.92& &$-27$&57&45.6&15.99&12002$\pm$38& &             &             &            12002$\pm$38\\
 20&58&57.37& &$-27$&56&01.8&19.04&12536$\pm$104& &             &             &            12536$\pm$104\\
 20&58&58.80& &$-28$&10&09.6&17.06& 9868$\pm$127& &             &             &             9868$\pm$127\\
 20&59&00.20& &$-28$&27&26.6&15.66& 9939$\pm$51& &             9931$\pm$13& &             9932$\pm$19\\
 20&59&00.53& &$-28$&15&22.5&14.98&11019$\pm$58& &            10978$\pm$16& &            10984$\pm$22\\
 20&59&01.54& &$-27$&55&49.8&18.71&12490$\pm$117& &             &             &            12490$\pm$117\\
 20&59&02.11& &$-28$&10&53.0&17.41&11621$\pm$35& &            11617$\pm$24& &            11619$\pm$25\\
 20&59&03.15& &$-28$&22&32.3&17.96&10133$\pm$91& &             &             &            10133$\pm$91\\
 20&59&03.66& &$-28$&35&44.8&15.12&10038$\pm$54& &            10031$\pm$14& &            10032$\pm$20\\
 20&59&04.96& &$-28$&04&06.2&15.42&12629$\pm$41& &             &             &            12629$\pm$41\\
 20&59&10.76& &$-27$&52&36.1&19.17&12696$\pm$57& &             &             &            12696$\pm$57\\
 20&59&12.23& &$-28$&38&46.7&17.62&10970$\pm$38& &             &             &            10970$\pm$38\\
 20&59&14.94& &$-27$&56&15.5&17.93&11335$\pm$37& &             &             &            11335$\pm$37\\
 20&59&16.90& &$-28$&14&32.2&     & &             &            11287$\pm$15& &            11287$\pm$23\\
 20&59&17.92& &$-28$&30&58.2&18.06& 9967$\pm$117& &             &             &             9967$\pm$117\\
 20&59&21.49& &$-28$&21&04.9&17.87&11574$\pm$83&11616$\pm$71& &             &            11598$\pm$54\\
 20&59&22.11& &$-28$&25&02.2&18.89&53255$\pm$130& &             &             &            53255$\pm$130\\
 20&59&22.21& &$-28$&34&52.7&18.31&10164$\pm$140& &             &             &            10164$\pm$140\\
 20&59&24.67& &$-28$&22&21.1&15.09& &            10919$\pm$38&10806$\pm$34&10933$\pm$41&10898$\pm$24\\
 20&59&24.77& &$-28$&18&51.4&18.79&10817$\pm$64& &             &             &            10817$\pm$64\\
 20&59&28.64& &$-27$&56&35.9&18.62&11415$\pm$54& &             &             &            11415$\pm$54\\
 20&59&31.53& &$-27$&53&17.0&16.89&12477$\pm$112& &             &             &            12477$\pm$112\\
 20&59&33.86& &$-28$&27&05.4&17.89&11091$\pm$34& &            11014$\pm$26& &            11058$\pm$26\\
 20&59&35.68& &$-28$&30&24.5&17.11&22740$\pm$69& &            22871$\pm$29& &            22833$\pm$37\\
 20&59&37.26& &$-28$&08&44.2&17.23&11722$\pm$31& &            11663$\pm$24& &            11697$\pm$23\\
 20&59&39.90& &$-28$&33&20.3&18.66&33012$\pm$103& &             &             &            33012$\pm$103\\
 20&59&43.65& &$-28$&42&29.5&17.87&12289$\pm$88& &             &             &            12290$\pm$88\\
 20&59&51.54& &$-28$&11&50.3&17.35&10574$\pm$40& &            10624$\pm$17& &            10609$\pm$22\\
 20&59&51.71& &$-28$&02&49.9&17.61&19646$\pm$68& &             &             &            19646$\pm$68\\
 20&59&52.12& &$-28$&16&46.7&16.74&10583$\pm$38& &            10628$\pm$16& &            10615$\pm$20\\
 20&59&55.83& &$-28$&27&00.0&17.82&26090$\pm$55& &             &             &            26090$\pm$55\\
 21&00&02.29& &$-28$&32&13.3&14.50&11864$\pm$46& &             &             &            11864$\pm$46\\
 21&00&04.11& &$-28$&04&41.2&18.49&12884$\pm$120&12678$\pm$58& &             &            12717$\pm$52\\
 21&00&04.30& &$-28$&19&06.0&17.97&11074$\pm$54& &            11093$\pm$21& &            11088$\pm$28\\
 21&00&04.40& &$-28$&07&09.0&19.15&57056$\pm$124& &             &             &            57056$\pm$124\\
 21&00&07.43& &$-27$&58&41.1&19.06&28971$\pm$151& &             &             &            28971$\pm$151\\
 21&00&11.51& &$-28$&38&20.5&18.65&11714$\pm$55& &             &             &            11714$\pm$55\\
 21&00&14.07& &$-28$&38&23.0&18.67&39535$\pm$139& &             &             &            39535$\pm$139\\
 21&00&24.06& &$-28$&32&28.6&18.72&11185$\pm$50& &             &             &            11185$\pm$50\\
 21&00&24.25& &$-28$&18&55.2&18.03&12287$\pm$33& &             &             &            12287$\pm$33\\
 21&00&26.31& &$-28$&32&51.1&16.65&11790$\pm$45& &             &             &            11790$\pm$45\\
 21&00&29.29& &$-28$&12&34.2&17.66&11144$\pm$129&11106$\pm$50& &             &            11111$\pm$47\\
 21&00&34.62& &$-28$&11&41.4&17.93&12080$\pm$39& &             &             &            12080$\pm$39\\
\end{tabular}
\end{flushleft}
\label{Table1cont}
\end{table*}

\setcounter{table}{0}
\begin{table*}
\caption[]{(cont.)}
\tablehead
 21&00&37.30& &$-28$&27&49.2&18.31&11838$\pm$92& &             &             &            11838$\pm$92\\
 21&00&40.51& &$-28$&07&34.5&18.51&26789$\pm$87& &             &             &            26789$\pm$87\\
 21&00&43.25& &$-28$&07&52.4&16.98&27033$\pm$64& &             &             &            27033$\pm$64\\
 21&00&44.08& &$-28$&14&44.9&18.47&12183$\pm$88& &             &             &            12183$\pm$88\\
 21&00&46.98& &$-28$&14&03.0&16.52& 9837$\pm$54& &             &             &             9837$\pm$54\\
 21&00&54.37& &$-28$&13&13.6&19.25& &            10619$\pm$50& &             &            10619$\pm$50\\
 21&00&55.91& &$-28$&13&06.9&18.35& &            10718$\pm$50& &             &            10718$\pm$50\\
 21&00&57.45& &$-28$&18&45.3&19.29&38749$\pm$72& &             &             &            38749$\pm$72\\
\noalign{\smallskip}
\hline
\end{tabular}
\end{flushleft}
\label{Table1contcont}
\end{table*}



\end{document}